\documentstyle[12pt,subeqnarray]{article}

\begin{document}

\title{Some Comments on \\ Multigrid Methods for Computing Propagators}

\author{
  \\
  {Alan D. Sokal}                   \\
  {\it Department of Physics}       \\
  {\it New York University}         \\
  {\it 4 Washington Place}          \\
  {\it New York, NY 10003 USA}      \\
  {\tt SOKAL@NYU.EDU}               \\
  \\
}
\vspace{0.5 cm}

\date{July 27, 1993}
\maketitle
\thispagestyle{empty}   

\newcommand{\be}{\begin{equation}}
\newcommand{\ee}{\end{equation}}
\newcommand{\<}{\langle}
\renewcommand{\>}{\rangle}
\newcommand{\widebar}{\overline}
\def\reff#1{(\protect\ref{#1})}
\def\spose#1{\hbox to 0pt{#1\hss}}
\def\ltapprox{\mathrel{\spose{\lower 3pt\hbox{$\mathchar"218$}}
 \raise 2.0pt\hbox{$\mathchar"13C$}}}
\def\gtapprox{\mathrel{\spose{\lower 3pt\hbox{$\mathchar"218$}}
 \raise 2.0pt\hbox{$\mathchar"13E$}}}
\def\textprime{${}^\prime$}
\def\proof{\par\medskip\noindent{\sc Proof.\ }}
\def\qed{\hbox{\hskip 6pt\vrule width6pt height7pt depth1pt \hskip1pt}}
\def\half{ {1 \over 2} }
\def\third{ {1 \over 3} }
\def\twothird{ {2 \over 3} }
\def\smfrac#1#2{\textstyle{#1\over #2}}
\newcommand{\imag}{\mathop{\rm Im}\nolimits}
\newcommand{\real}{\mathop{\rm Re}\nolimits}
\newcommand{\Ran}{\mathop{\rm Ran\,}\nolimits}
\def\dslash{ {D \!\!\!\!\!\: / \,} }
\def\hboxscript#1{ {\hbox{\scriptsize\em #1}} }
\def\Heff#1{ {H_{#1}^{\hbox{\scriptsize\em eff}}} }

\def\scra{{\cal A}}
\def\scrc{{\cal C}}
\def\scrf{{\cal F}}
\def\scrl{{\cal L}}
\def\scrm{{\cal M}}
\def\scro{{\cal O}}
\def\scrr{{\cal R}}
\def\scrs{{\cal S}}
\def\scrt{{\cal T}}
\def\scrv{{\cal V}}
\def\scrz{{\cal Z}}

\def\r{{\bf r}}
\def\e{{\bf e}}
\def\q{{\bf q}}
\def\k{{\bf k}}
\def\fN{ {f_{\hbox{\scriptsize N}}} }
\def\fW{ {f_{\hbox{\scriptsize W}}} }
\def\fZ{ {f_{\hbox{\scriptsize Z}}} }
\def\MbarN{ {{\bar M}_{\hbox{\scriptsize N}}} }
\def\MbarW{ {{\bar M}_{\hbox{\scriptsize W}}} }
\def\MbarZ{ {{\bar M}_{\hbox{\scriptsize Z}}} }
\def\var{ {\hbox{var}} }
\def\cov{ {\hbox{cov}} }
\def\Cov{ {\hbox{Cov}} }

\def\gtil{ {\widetilde{g}} }

\newtheorem{theorem}{Theorem}[section]
\newtheorem{proposition}[theorem]{Proposition}
\newtheorem{lemma}[theorem]{Lemma}
\newtheorem{corollary}[theorem]{Corollary}

%
%
\newenvironment{sarray}{
          \textfont0=\scriptfont0
          \scriptfont0=\scriptscriptfont0
          \textfont1=\scriptfont1
          \scriptfont1=\scriptscriptfont1
          \textfont2=\scriptfont2
          \scriptfont2=\scriptscriptfont2
          \textfont3=\scriptfont3
          \scriptfont3=\scriptscriptfont3
        \renewcommand{\arraystretch}{0.7}
        \begin{array}{l}}{\end{array}}

\newenvironment{scarray}{
          \textfont0=\scriptfont0
          \scriptfont0=\scriptscriptfont0
          \textfont1=\scriptfont1
          \scriptfont1=\scriptscriptfont1
          \textfont2=\scriptfont2
          \scriptfont2=\scriptscriptfont2
          \textfont3=\scriptfont3
          \scriptfont3=\scriptscriptfont3
        \renewcommand{\arraystretch}{0.7}
        \begin{array}{c}}{\end{array}}

%
%
%
\font\srm=cmr7                             
\font\tenmsx=msxm10 scaled\magstep1        
\font\specialroman=msym10 scaled\magstep1  
\font\sevenspecialroman=msym7              

%
%
\def\zed{{\hbox{\specialroman Z}}}
\def\szed{{\hbox{\sevenspecialroman Z}}}
\def\R{{\hbox{\specialroman R}}}
\def\sR{{\hbox{\sevenspecialroman R}}}
\def\C{{\hbox{\specialroman C}}}
\def\sC{{\hbox{\sevenspecialroman C}}}
\renewcommand{\emptyset}{{\hbox{\specialroman ?}}}
\def\restrict{{\hbox{\tenmsx \char"016}}}

\begin{abstract}
I make three conceptual points regarding multigrid methods for
computing propagators in lattice gauge theory:
1)  The class of operators handled by the algorithm must be stable
under coarsening.
2)  Problems related by symmetry should have solution methods related by
symmetry.
3)  It is crucial to distinguish the vector space $V$ from its dual space
$V^*$.
All the existing algorithms violate one or more of these principles.
\end{abstract}

\vspace{0.5 cm}

\clearpage

There has recently been much interest in developing multigrid methods for
solving large linear systems with disordered coefficients
\cite{EGS_PRL,EGS_LAT88,Edwards_thesis},
and in particular for computing the bosonic or fermionic propagator
in a background gauge field.\footnote{
   Typically the gauge field $U$ is chosen stochastically from some
   probability distribution $P(U)$, e.g.\ either that of quenched
   lattice gauge theory or else that of lattice gauge theory with
   dynamical fermions.
}
Many interesting ideas in this direction have been proposed by
the Amsterdam \cite{Hulsebos_LAT88,Hulsebos_90,Hulsebos_LAT90,Vink_91,%
Hulsebos_92a,Hulsebos_92b,Vink_LAT91},
Boston \cite{Brower_91a,Brower_91b,Brower_91c,Brower_91d,Brower_LAT90},
Hamburg \cite{Mack_88,Kalkreuter_92a,Kalkreuter_92b,Kalkreuter_92c,%
Kalkreuter_92d,Kalkreuter_LAT92,Baker_93,Baker_LAT92,Kalkreuter_93a,%
Kalkreuter_93b}
and Israeli \cite{BenAv_90,BenAv_91,Harmatz_LAT90,BenAv_thesis,Lauwers_92a,%
Lauwers_92b,Harmatz_LAT92,Lauwers_LAT92,Lauwers_93,BenAv_93}
groups.
However, all these discussions have missed what I consider to be
some key conceptual points.
Perhaps a brief note explaining these points is therefore warranted.

\bigskip

{\em First point.}\/
We are interested in solving linear systems of the form $Ax=b$
for some class of linear operators $A$.
So the first order of business must be to specify clearly
the class $\scrc$ of operators to which our algorithm is intended to apply.
This is not a completely trivial matter, because
\begin{itemize}
   \item[1)]  For a {\em multigrid}\/ (as opposed to two-grid) algorithm
      to be well-defined,
      the class $\scrc$ must be stable under coarsening.\footnote{
  I emphasize that the issue here is not whether the algorithm
  performs well or poorly;  the issue is whether the algorithm
  is meaningful at all.
}
\end{itemize}
In other words, we start with some class $\scrc_0$ of operators on
the finest grid, for which we want to solve $Ax=b$.
But then our multigrid algorithm produces operators on the first coarse grid,
which belong to some {\em quite possibly larger}\/ class $\scrc_1$
--- what this class is depends on our procedure for choosing interpolation,
restriction and coarse-grid operators.
If we are then to continue to the {\em second}\/ coarse grid,
this procedure
must be defined for all the operators
in this larger class.  If we want to allow an arbitrary number of grids,
then we cannot stop short of a class $\scrc \supset \scrc_0$
that is {\em stable under coarsening}\/.

For example, our initial interest might be in nearest-neighbor
Laplace operators in an $SU(N)$ gauge field, with $x$-independent
values for the hopping parameter $\beta$ and the mass $m$:
\be
   \scrc_0   \;=\;
   \{ A \equiv -\beta\Delta_U + m^2 \colon\;
      U_{xy} \in SU(N), \, \beta \ge 0, \, m^2 \in \R, \, A > 0 \}
   \;,
 \label{eq1}
\ee
where
\be
   (\Delta_U)_{xy}   \;=\;
   \cases{  -2d    & for $x=y$   \cr
            U_{xy} & for $|x-y| = 1$  \cr
            0      & otherwise \cr
         }
\ee
But quite possibly our coarsening procedure will produce a {\em larger}\/
class of operators on the coarse grids:
the mass may vary from site to site, or may even become non-trivial
in color space;
and the off-diagonal elements may have magnitudes that vary from link to link,
or they may even leave the space $\R_+ \times SU(N)$ and enter its
linear span (which for $N \ge 3$ is the space of all $N \times N$
complex matrices).\footnote{
   This is a ``dielectric gauge field'' in the language of Mack
   \cite{Mack_dielectric}.
}
If this occurs, we must then go back to the beginning, and define our
coarsening procedure for all the operators in this larger class ---
and so on successively until the class stabilizes.\footnote{
   This criterion is irrelevant in {\em unigrid}\/
   \cite{McCormick-Ruge_unigrid} approaches
   (as proposed e.g.\ in \cite{Baker_93,Baker_LAT92}),
   since here the concept of a coarse-grid operator never arises:
   everything is done on the original (fine) grid.
   However, these approaches have a severe drawback:
   the interpolation kernel must be computed on arbitrarily big blocks
   (e.g. $2^k$ or $3^k$ for arbitrarily large $k$),
   and this can be prohibitive if the CPU time for this computation
   grows faster than linearly in the block volume.
}

This comment is relevant to the ground-state-projection (GSP)
multigrid algorithms proposed by the
Boston \cite{Brower_91a,Brower_91b,Brower_91c,Brower_91d,Brower_LAT90}
and Hamburg \cite{Kalkreuter_92a,Kalkreuter_92b,Kalkreuter_92c,%
Kalkreuter_92d,Kalkreuter_LAT92,Kalkreuter_93a,%
Kalkreuter_93b}
groups,
in which the range of the interpolation operator is spanned by
the lowest eigenvectors of the operator $A$ modified to impose
Neumann boundary conditions on $2 \times 2$ or $3 \times 3$ blocks.
The trouble here is that the concept of ``Neumann b.c.'',
though well-defined for operators of the class \reff{eq1},
has no unambiguous meaning (as far as I can tell)
when the off-diagonal elements of $A$ are more general than
(positive real number) $\times$ (unitary matrix).
Possibly an appropriate definition can be found,
but this is a highly nontrivial problem,
particularly when the required covariances are taken into account
(see the ``second point'' below).
The situation is different for the
Amsterdam version of GSP multigrid
\cite{Hulsebos_LAT88,Hulsebos_90,Hulsebos_LAT90,Hulsebos_92a,%
Hulsebos_92b}
(see also \cite{Kalkreuter_92a}),
which employs {\em Dirichlet}\/ b.c.\ on $2 \times 2$ blocks.
This operator {\em is}\/ well-defined in general:
one simply sets to zero those off-diagonal elements of $A$
that cross block boundaries.
On the other hand, it is unclear whether Dirichlet b.c.\ give
a {\em good}\/ interpolation:
for example, in the case of the pure Laplacian ($U \equiv I$),
Dirichlet b.c.\ give a reasonable interpolation only on $2 \times 2$ blocks,
where they are {\em equivalent}\/
to Neumann b.c.\ plus a multiple of the identity matrix
(and yield piecewise-constant interpolation);
on larger blocks, Dirichlet b.c.\ yield sine-wave interpolation kernels,
which do {\em not}\/ perform well (because smooth functions,
e.g.\ constants, cannot be interpolated accurately).
This is not necessarily an argument against
Dirichlet b.c.\ on $2 \times 2$ blocks,
but it does raise doubts about whether the {\em concept}\/
behind GSP using Dirichlet b.c.\ is correct.

\bigskip

{\em Second point.}\/
Suppose that there is a transformation $T\colon\, \scrc \to \scrc$
that is ``trivially calculable'',
in the sense that from the solution of $Ax=b$ we can trivially determine
(in a negligible CPU time)
the solution of $T(A) x = b'$, and conversely.
Then any sensible method (whether multigrid or anything else)
for solving problems in $\scrc$ ought to be {\em covariant}\/ under $T$,
in the sense that using the method to solve $Ax=b$ ought to give exactly
the same approximate solutions, in exactly the same CPU time,
as transforming to $T(A) x = b'$, using the method to solve this latter
equation, and then transforming back.
Why?  Well, suppose it weren't so.
Then we would have {\em two alternative}\/ methods for solving $Ax=b$
and $T(A) x = b'$, and presumably one of them would be {\em better}\/
than the other;  we should then use {\em this}\/ method for {\em both}\/
equations, and discard the other method.
The principle here is quite general, and has nothing to do {\em per se}\/
with linear equations:
\begin{itemize}
   \item[2)]  Problems related by symmetry should have solution methods
      related by symmetry.
      (A special case is:  Symmetric problems should have symmetric
      solution methods.)
\end{itemize}
Of course, this principle is not {\em always}\/ valid:
there could arise, in principle, cases of
``spontaneous symmetry breaking in algorithm space'',
in which each of a family of non-$T$-covariant algorithms would be better
than the best $T$-covariant algorithm.  But I find this possibility
extremely unlikely in the present context.

This principle has powerful consequences in combination with the first point.
For example, the class $\scrc_0$ defined above is mapped into itself
under $SU(N)$ gauge transformations and under {\em space-independent}\/
rescalings $A \to \lambda A$;
and many workers have recognized that a good algorithm ought to be covariant
under these transformations.
But the larger class of operators produced by coarsening may well be mapped
into itself by {\em space-dependent}\/ rescalings
$A_{xy} \to \lambda_x A_{xy} \lambda_y^*$
or even by $GL(N,\C)$ gauge transformations
$A_{xy} \to U_x A_{xy} U_y^*$ with $U_x \in GL(N,\C)$.
The multigrid algorithm ought then to be covariant under this larger group;
but none of the published algorithms, to my knowledge,
have this property.\footnote{
   The parallel-transported multigrid (PTMG) algorithm proposed by the
   Israeli group
   \cite{BenAv_90,BenAv_91,Harmatz_LAT90,BenAv_thesis,Lauwers_92a,Lauwers_92b,%
Harmatz_LAT92,Lauwers_LAT92,Lauwers_93,BenAv_93}
   avoids this problem by {\em forcing}\/ the coarse-grid operator
   to lie in the original class $\scrc_0$.
   The price paid is that the coarse-grid operator does not
   satisfy the Galerkin condition $A_1 = R A_0 P$
   (see footnote \ref{footnote_Galerkin} below for discussion of
    why the Galerkin condition is desirable).
    The divergent iteration observed at small fermion mass
    \cite{BenAv_91,Harmatz_LAT90,BenAv_thesis,Lauwers_92a,Lauwers_93,BenAv_93}
    is probably related to this choice of a non-Galerkin coarse-grid operator.
}

\bigskip

{\em Third point.}\/
My third point is more abstract, but it is related to the second one.
Consider again $Ax=b$.  On both physical and mathematical grounds,
\begin{itemize}
   \item[3)]  It is crucial to distinguish between the vector space $V$
      of solutions $x$ and the vector space $W$
      of ``forcing terms'' $b$.\footnote{
  Two vector spaces of the same finite dimension are of course isomorphic,
  but they are not {\em naturally}\/ isomorphic.  Otherwise put,
  there are {\em many}\/ linear bijections from $V$ to $W$,
  and if we want to single out one of them we must find a {\em physical}\/
  reason for doing so.
}
\end{itemize}
Thus, $x$ may be a displacement while $b$ is a force;
$x$ may be a magnetization while $b$ is a magnetic field; and so forth.
Most commonly in physical applications, we have $W = V^* =$ dual space of $V$:
that is, there is a {\em natural}\/ nondegenerate bilinear mapping
$V \times W \to \R$, which typically can be interpreted as an energy
(force $\cdot$ displacement,
magnetic field $\cdot$ magnetization, etc.).\footnote{
   I use the notations $\ell(x) \equiv \< \ell, x \> \equiv \< x, \ell \>$
   to denote the natural bilinear mapping acting on
   $\ell \in V^*$ and $x \in V$.
}
In this case the principle becomes
\begin{itemize}
  \item[3\textprime)]  It is crucial to distinguish between the
      vector space $V$ and its dual space $V^*$.
      (In a language perhaps more familiar to physicists:
       it is crucial to distinguish between contravariant and covariant
       tensors.)
\end{itemize}
This principle is well known to differential geometers and general relativists,
who have learned (since roughly the 1950's and 1960's, respectively)
the advantages of a ``coordinate-free'' (basis-independent) notation.
However, this principle is often forgotten by elementary-particle physicists,
because we typically learned our linear algebra in the context of either
quantum mechanics or matrix theory --- each of which works {\em not}\/
on a bare vector space $V$, but rather on a vector space $V$
{\em equipped with additional structure}\/:

(a) Quantum mechanics employs a {\em Hilbert space}\/, i.e.\ a vector space
equipped with an {\em inner product}\/
(positive-definite bilinear form on $V$).
This inner product induces an identification map between $V$ and $V^*$.
In quantum mechanics the inner product has, of course, a good physical
interpretation (transition amplitude).  In other contexts we should
introduce such an identification operator {\em only}\/ if it has likewise
a good physical or mathematical {\em raison d'\^etre}\/.

(b) Matrix theory is linear algebra {\em referred to a particular basis}\/
--- and this can be very misleading, unless the basis has been selected
by some good physical or mathematical criterion.
For example, if we have chosen particular bases for $V$ and $V^*$,
then the matrix with entries $\delta_{ij}$ defines a particular
identification map from $V$ to $V^*$;  but this identification is {\em not}\/
natural, i.e.\ it is of no more physical or mathematical interest than any
other linear bijection from $V$ to $V^*$.
(In different bases this same identification would have {\em different}\/
 matrix representations, {\em not}\/ equal to $\delta_{ij}$.)
In particular, it is misleading to employ the term ``identity matrix''
in this context;  this term ought to be reserved for the matrix representation
of the {\em identity operator}\/, which maps a space $V$ into {\em itself}\/.
(This operator has the matrix representation $\delta_{ij}$
with respect to {\em any}\/ basis.)

As soon as we distinguish between $V$ and $V^*$, we must distinguish
two very different types of linear operators:
\begin{itemize}
  \item  Linear operators $A \colon\, V \to V$.
  \item  Linear operators $Q \colon\, V \to V^*$  [or what is equivalent,
     bilinear forms $\widehat{Q} \colon\, V \times V \to \R$].\footnote{
 The correspondence is $\widehat{Q}(x,y) = (Qx)(y)$ for $x,y \in V$.
}${}^,$\footnote{
 In the complex case one has to consider {\em sesqui}\/linear forms,
 and the role of $V^*$ is played by the {\em anti}\/dual space
 (i.e.\ the space of all {\em anti}\/linear maps from $V$ to $\C$).
 For simplicity we shall henceforth restrict attention to the real case.
 This is no loss of generality, because a complex vector space of
 dimension $n$ can be considered as a real space of dimension $2n$;
 and under this correspondence eigenvalues are preserved
 (more precisely, $a+bi$ becomes the {\em pair}\/ $a \pm bi$),
 hermiticity corresponds to symmetry, etc.
}
\end{itemize}
Only for operators from $V$ to itself can we discuss eigenvalues,
 spectral radius, trace, determinant, etc.;
only for operators from $V$ to $V^*$ can we discuss symmetry (or hermiticity)
 and positive-(semi)definiteness.\footnote{
   Symmetry means that $(Qx)(y) = (Qy)(x)$, or equivalently that
   $\widehat{Q}(x,y) = \widehat{Q}(y,x)$.
   Symmetry can also be written as $Q = Q^*$, once we remember the
   definition of dual operator:
   if $A\colon\, V \to W$, then $A^* \colon\, W^* \to V^*$
   is defined by $(A^* x)(y) = x(Ay)$ for $x \in W^*$, $y \in V$.
}${}^,$\footnote{
   Positive-semidefiniteness means that $(Qx)(x) \ge 0$ for all $x \in V$.
   Positive-definiteness means that $(Qx)(x) > 0$ for all $x \in V$,
   $x \neq 0$.
}

Thus, in most physical contexts the Laplacian arises naturally as an
operator {\em from $V$ to $V^*$}\/:
the corresponding quadratic form is the energy functional.
It is thus {\em meaningless}\/ to discuss the ``eigenvalues'' of such
an operator.  (What is actually meant by such discussions will be
explained below.)

Let us examine the multigrid method in this context.  We are given
a fine-grid operator $A_0 \colon\, V_0 \to V_0^*$
--- usually symmetric and positive-definite ---
for which we want to solve $A_0 x=b$.
We then introduce a coarse-grid space $V_1$, along with
restriction, interpolation (prolongation) and coarse-grid operators.
On what spaces do these operators act?
Examination of the multigrid algorithm indicates that restriction
is always applied to a {\em residual}\/ (right-hand side);
thus, $R \colon\, V_0^* \to V_1^*$.
Similarly, interpolation is always applied to an {\em error}\/
(left-hand side);
thus, $P \colon\, V_1 \to V_0$.
Finally, the coarse-grid operator acts similarly to the fine-grid operator,
i.e.\ $A_1 \colon\, V_1 \to V_1^*$.
We thus have the following diagram:
%
\def\mapdown#1{\Big\downarrow\rlap{$\vcenter{\hbox{$\scriptstyle#1$}}$}}
\def\mapup#1{\llap{$\vcenter{\hbox{$\scriptstyle#1$}}$}\Big\uparrow}
\be
\begin{array}{ccc}
   V_0         &  \stackrel{A_0}{\longrightarrow}   &   V_0^*   \\
   \mapup{P}   &                                    &  \mapdown{R}  \\
   V_1         &  \stackrel{A_1}{\longrightarrow}   &   V_1^*
\end{array}
\ee
It follows that the oft-imposed condition $R = P^*$ is {\em meaningful}\/.
Likewise, the Galerkin condition $A_1 = R A_0 P$
is {\em meaningful}\/.\footnote{
   \label{footnote_Galerkin}
   Briggs \cite[pp.~67--69]{Briggs_87}
   and the Hamburg group \cite{Kalkreuter_92c,Baker_93,Baker_LAT92}
   have given a very strong
   argument for the Galerkin choice of the coarse-grid operator.
   After fine-grid relaxation, we have an error $e_0 \equiv x_0 - A_0^{-1} b$
   which satisfies the equation $A_0 e_0 = r_0$, where $r_0 \equiv A_0 x_0 -b$
   is the residual.  Now, the multigrid philosophy is based on the idea
   that this error is {\em smooth}\/, in the sense that it is near to
   the range of the interpolation operator, i.e.\ $e_0 \approx P e_1$
   for some vector $e_1 \in V_1$.  This vector $e_1$ satisfies the equation
   $R A_0 P e_1 \approx r_1$, where $r_1 \equiv R r_0$ is the coarse-grained
   residual.  So {\em this}\/ is the equation we should aim to solve
   on the coarse grid, i.e.\ we should take $A_1 = R A_0 P$.
   (Note that this derivation does {\em not}\/ need the condition $RP = I$
    \cite{Kalkreuter_92c,Baker_93,Baker_LAT92},
   which in fact is meaningless.)
   This derivation holds whether or not $A_0$ is symmetric and/or
   positive-definite, and whether or not $R = P^*$.

   A slight variant of this argument, using the $=$ sign instead of $\approx$,
   goes as follows:
   In the coarse-grid-correction phase of the multigrid algorithm,
   we replace $x_0$ by $x_0 +Px_1$ for some $x_1 \in V_1$.
   How should we choose $x_1$?
   Ideally we would like the new vector $x_0 +Px_1$ to have a
   {\em zero}\/ residual, i.e.\ $A_0 (x_0 +Px_1) - b = 0$;
   but this is obviously too much to hope for
   in general, since $\dim V_1 < \dim V_0$.
   The best we can hope for is to annihilate the {\em smooth part}\/
   of the residual, i.e.\ to have $R [A_0 (x_0 +Px_1) - b] = 0$.
   (As before, we expect that the {\em non-smooth}\/ part of the residual
    has already been made small by the fine-grid relaxation.)
   We should therefore aim to choose $x_1$ so that $RA_0 P x_1 = -r_1$.
}${}^,$\footnote{
   If the fine-grid operator $A_0$ is symmetric
   (resp.\ symmetric and positive-semidefinite)
   and we make the variational choices $R = P^*$ and $A_1 = R A_0 P$,
   then the coarse-grid operator $A_1$ is likewise symmetric
   (resp.\ symmetric and positive-semidefinite) ---
   a useful property.
}${}^,$\footnote{
   \label{footnote_variational}
   If the fine-grid operator $A_0$ is symmetric and positive-definite,
   then we can give a strong argument
   \cite[pp.~56--57]{Brandt_84}
   \cite[p.~2043]{Goodman-Sokal_MGMC1}
   for the variational choices $R = P^*$ and $A_1 = R A_0 P$.
   Note first that solving $A_0 x = b$ is equivalent
   to minimizing the energy functional
   $H_0(x) = {1 \over 2} \<x, A_0 x\> - \< b,x \>$.
   Now, in the coarse-grid-correction phase of the multigrid algorithm,
   we replace the current approximate solution $x_0$ by a hopefully better
   approximate solution $x_0 + P x_1$, where $x_1 \in V_1$.
   A sensible goal is to attempt to choose $x_1$ so as to minimize $H_0$;
   this means that we attempt to minimize
   $H_1(x_1) \equiv H_0(x_0 + P x_1) =
        {1 \over 2} \<x_1, P^* A_0 P x_1\> - \< P^* (b- A_0 x_0), x_1 \>
        + {\rm const}$.
   This is precisely what the multigrid algorithm with $R = P^*$ and
   $A_1 = P^* A_0 P$ does.
   (The only loophole in this logic is that one might entertain the
    possibility of ``coarse-grid overcorrection''
    \cite{Meis_82,Goodman-Reyna_private,Kalkreuter_92b,Kalkreuter_93a},
    i.e.\ of choosing temporarily a {\em sub-optimal}\/ approximate solution
    $x_1$ in the hope that this will later bring benefits.)
}
On the other hand, the condition $RP = I$
\cite{Mack_88,Kalkreuter_92a,Kalkreuter_92c,Baker_93,Baker_LAT92}
is {\em meaningless}\/.

However, this is not the end of the story, because the multigrid algorithm
contains an additional ingredient, namely a relaxation (smoothing) operator
$\scrs_0 \colon\, V_0 \times V_0^* \to V_0$
which takes an approximate solution of $A_0 x = b$ and produces a new
(and hopefully better) approximate solution.
Usually this relaxation is a first-order stationary linear process
\be
   \scrs_0(x,b)   \;=\;   S_0 x  +  T_0 b   \;,
 \label{star1}
\ee
where $S_0 \colon\, V_0 \to V_0$, $\, T_0 \colon\, V_0^* \to V_0$
and the consistency condition is
$T_0 = (I-S_0) A_0^{-1}$ [verify that this formula is meaningful!].
Such a process ordinarily arises from a {\em splitting}\/
\be
  A_0   \;=\;   B_0 + C_0   \;,
 \label{star2}
\ee
where $B_0 \colon\, V_0 \to V_0^*$ is some operator that is
``easy to invert'';
writing $A_0 x = b$ as $B_0 x = -C_0 x + b$ suggests the iteration
\reff{star1} with
\begin{subeqnarray}
   S_0   & = &  - B_0^{-1} C_0   \;=\;   I  - B_0^{-1} A_0    \\
   T_0   & = &  B_0^{-1}
\end{subeqnarray}
[Conversely, any convergent first-order stationary linear iteration
\reff{star1} arises from a splitting $B_0 = T_0^{-1} = A_0 (I - S_0)^{-1}$.]
A ``good'' choice of $B_0$ (for a {\em single-grid}\/ algorithm)
is one that is ``as close to $A_0$ as possible'', subject to the
condition that it be ``easy to invert'':
this makes $S_0$ as small as possible.

Let us now examine the oft-made statement that
``the slow modes of the (damped) Jacobi iteration correspond to the
low eigenvalues of $A_0$''.
This statement is in fact {\em meaningless}\/,
because $A_0$ does not map a space into itself,
and thus does not have eigenvalues.
But it is easy to see what the correct statement is:
the slow modes of \reff{star1} correspond to the eigenvalues near 1 of $S_0$,
hence to the low eigenvalues of $B_0^{-1} A_0$.
Here $B_0^{-1} A_0$ maps $V_0$ into itself, and so {\em does}\/ have
eigenvalues.

But we can also understand how the meaningless statement could have arisen.
The (damped) Jacobi iteration is defined by letting $B_0$ be
(a multiple of) the diagonal part of $A_0$
{\em with respect to some particular basis}\/.
Now, for certain operators $A_0$ in certain bases
--- for example, the Laplace operator $-\beta\Delta_U + m^2$
in a site-wise basis ---
it happens that the diagonal part of $A_0$ is given by a multiple
of the identity matrix, so it is easy to {\em think}\/ (incorrectly)
that we are discussing the eigenvalues of $A_0$ rather than those of
$B_0^{-1} A_0$.
I emphasize that the performance of the Jacobi algorithm is a property
not only of $A_0$ but also of the choice of basis:
for example, for the {\em same}\/ Laplace operator $A_0 = -\beta\Delta_U + m^2$
with $U \equiv I$,
if we use a Fourier basis we have $B_0 = A_0$,
in which case the relaxation process has {\em no}\/ slow modes.
On the other hand, for more general operators $A_0$
--- such as Laplace operators with a site-dependent mass term ---
the diagonal part of $A_0$ is {\em not}\/ a multiple of the identity matrix,
even in the site-wise basis;
so in such a case the conventional statement is not merely
misleading but is in fact {\em incorrect}\/.
In all cases, the relevant quantities are the eigenvalues and eigenvectors
of $B_0^{-1} A_0$.

Returning to the multigrid algorithm, we can summarize the situation
as follows:
\begin{itemize}
   \item[i)]  The fine-grid problem is specified by an operator
      $A_0 \colon\, V_0 \to V_0^*$,
      which is usually symmetric and positive-definite
      (as will be assumed henceforth).
      {\em No}\/ other structure is present in the original problem.
   \item[ii)]  Choosing a fine-grid relaxation method amounts to choosing
      a second operator $B_0 \colon\, V_0 \to V_0^*$.
      This operator is often (but not always) symmetric and
       positive-definite.\footnote{
   The (damped) Jacobi choice of $B_0$ is symmetric and
   positive-definite.  The Gauss-Seidel or SOR choice of $B_0$ is
   usually {\em not}\/ symmetric, but it may be made symmetric
   (as well as positive-semidefinite) by explicit symmetrization,
   e.g.\ a forward sweep followed by a reverse sweep.
   Note also that $B_0$ is symmetric if and only if
   $S_0 \equiv I - B_0^{-1} A_0$ is $A_0$-symmetric
   (this means that $A_0 S_0 = S_0^* A_0$).
}
\end{itemize}
We must then choose the remaining elements of the multigrid algorithm:
$R$, $P$ and $A_1$.  Obviously these choices can and should depend on
$A_0$ and $B_0$;  but they should {\em not}\/ depend on any {\em additional}\/
structures unless we have a good physical or mathematical reason for
introducing such structures into the problem.

In footnotes \ref{footnote_Galerkin}--\ref{footnote_variational} above,
I have given strong arguments for the variational choices
$R = P^*$ and $A_1 = P^* A_0 P$.
Henceforth these choices will be assumed.
The only remaining choice is therefore that of the interpolation operator $P$.
(This is, of course, the heart of the problem is disordered systems.)

The goal in the choice of $P$ is obviously to maximize the speed of convergence
of the algorithm, measured in CPU-time units.
Now, the asymptotic rate of convergence of any
first-order stationary linear iteration
is $-\log \rho(M)$, where $\rho(M)$ is the spectral radius\footnote{
   The spectral radius of an operator $M \colon\, V \to V$
   is the largest absolute value of an eigenvalue.
}
of the algorithm's iteration matrix $M \colon\, V \to V$.
The iteration matrix of the multigrid algorithm is rather complicated ---
it depends on the number of levels, on the choice of V-cycle or W-cycle,
etc.\ ---
so it is convenient to consider instead the iteration matrix of the
{\em idealized two-grid algorithm}\/,
i.e.\ the two-grid algorithm in which one assumes that the coarse-grid
equation $A_1 x_1 = b_1$ is solved {\em exactly}\/.
This is a useful warm-up problem prior to studying the complete multigrid
algorithm, and there are both heuristic and rigorous connections between
the two algorithms.\footnote{
 \label{footnote_energy_norm}
   Obviously the good performance of the idealized two-grid algorithm is a
   {\em necessary}\/ condition for the good performance of the multigrid
   algorithm.  On the other hand, it is also a {\em sufficient}\/ condition,
   at least for the W-cycle:  if the energy norm of the
   idealized-two-grid iteration matrix is
   $\le (\sqrt{2}/2) - \epsilon$
   uniformly in the lattice size, then the energy norm of the
   multigrid iteration matrix is $\le 1 -\delta$
   uniformly in the lattice size \cite{McCormick_82,Maitre_84}.
   (The energy norm of an operator $M \colon\, V_0 \to V_0$
    is $\| M \| = \sup\limits_{x \in V_0}
        ( \< Mx, A_0 Mx \>  /  \< x, A_0 x \> )^{1/2}$.)
}
Let us therefore concentrate henceforth on the iteration matrix of the
idealized two-grid algorithm, for simplicity with only pre-smoothing:
\be
   M_0   \;=\;   (I - P A_1^{-1} R A_0) S_0  \;.
\ee
It is worth noting that the operator $P A_1^{-1} R A_0 \colon\, V_0 \to V_0$
is the $A_0$-orthogonal projection onto the range of $P$.\footnote{
   {\em First proof (computational):}\/
   Clearly $\Ran  P A_1^{-1} R A_0 \subset  \Ran P$.
   To show that $P A_1^{-1} R A_0$ acts as the identity on $\Ran P$,
   let us apply it to a vector $x_0 = P x_1$ where $x_1 \in V_1$:
   we get $P A_1^{-1} R A_0 P x_1 = P x_1$ since $R A_0 P = A_1$.
   Finally, we note that $P A_1^{-1} R A_0$ is $A_0$-symmetric,
   i.e.\ $A_0 (P A_1^{-1} R A_0) = (P A_1^{-1} R A_0)^* A_0$;
   this is an immediate consequence of $A_0^* = A_0$, $A_1^* = A_1$ and
   $R = P^*$.

   {\em Second proof (conceptual):}\/
   By construction, the coarse-grid-correction phase of the idealized two-grid
   algorithm with the variational choices $R = P^*$ and $A_1 = P^* A_0 P$
   replaces the vector $x_0$ by that vector $x_0 +Px_1$ ($x_1 \in V_1$)
   which minimizes the energy
   ${1 \over 2} \< (x_0 +Px_1), A_0 (x_0 +Px_1) \>$.
   (Here we can assume without loss of generality a right-hand side $b=0$.)
   This means that the operator $I - P A_1^{-1} R A_0$
   is the $A_0$-orthogonal projection onto the $A_0$-orthogonal complement
   of $\Ran P$.
}

The condition for minimizing the spectral radius of $M_0$ was given
some years ago by Greenbaum \cite{Greenbaum_84}:
If $\dim V_1 = n_1$, then $\rho(M_0)$ is minimized when
$\Ran P$ consists of the $n_1$ eigenvectors of $B_0^{-1} A_0$
with the smallest eigenvalues.\footnote{
   Actually, Greenbaum's condition \cite{Greenbaum_84}
   is slightly different, as she seeks to minimize the {\em energy norm}\/
   of $M_0$ (see footnote \ref{footnote_energy_norm} for its definition)
   rather than the spectral radius.
   The two criteria are equivalent whenever $B_0$ is symmetric
   (i.e.\ $S_0$ is $A_0$-symmetric).
}
Note that only the {\em range}\/ of the interpolation is constrained here:
we can compose $P$ with an arbitrary bijective map in $V_1$,
and as long as we use the variational choices $R = P^*$ and $A_1 = P^* A_0 P$,
this merely amounts to a redefinition of the coarse-grid fields,
which has no effect on the sequence of approximants $x_0$ that the
algorithm computes.

To be sure, Greenbaum's condition is ``impractical'', for two reasons:
Firstly, it is usually unfeasible to compute the eigenvectors of $B_0^{-1} A_0$
(unless $A_0$ and $B_0$ are constant-coefficient difference operators),
as this would take a CPU time of order (volume)${}^3$.
Secondly, even if we could compute the eigenvectors of $B_0^{-1} A_0$,
we would not want to use them as the columns of the interpolation matrix,
because the matrices $P$, $R$ and $A_1$ would then be {\em dense}\/ rather
than sparse, and each iteration of the resulting multigrid algorithm
would take a CPU time of order (volume)${}^2$.
Nevertheless, Greenbaum's condition is important because it indicates
the {\em ideal}\/ towards which we are striving,
even though in practice this ideal will have to be traded off
against considerations of computational complexity.

It is in this context that we can consider
the Hamburg group's recent proposal \cite{Baker_93,Baker_LAT92},
which goes as follows:
The ``best'' idealized two-grid algorithm is the one with the smallest
spectral radius of $M_0$.
But spectral radii are difficult to compute,
so let us study instead some {\em norm}\/ of $M_0$;
this changes the problem, but hopefully not by too much.
Now, norms are often difficult to compute, too;
but one norm which is easy to compute is the
Frobenius (Hilbert-Schmidt) norm
$\| M_0 \| _2 ^2   =  {\rm tr}\, M_0 M_0^* $.
The first trouble --- as the reader should by now be able to guess ---
is that this expression is {\em meaningless}\/:
the range of $M_0^*$ is in $V_0^*$, while the domain of $M_0$ is $V_0$.
However, we can fix it by defining instead\footnote{
   A similar use of $B_0$ --- or more precisely, of the diagonal part of $A_0$
   --- can be found in the norms defined by Ruge and St\"uben
   \cite[p.~78]{Ruge_87}.
}
\begin{eqnarray}
   \| M_0 \| _2 ^2  & = &  {\rm tr}\, B_0 M_0 B_0^{-1} M_0^*      \nonumber \\
                & (=\hphantom{)} &
      {\rm tr}\, M_0 B_0^{-1} M_0^* B_0  \;=\;
      {\rm tr}\, B_0^{-1} M_0^* B_0 M_0  \;=\;
      {\rm tr}\, M_0^* B_0 M_0 B_0^{-1} )  \;.
  \label{trace_norm}
\end{eqnarray}
This expression is meaningful without regard to any choice of basis,
and it agrees with the preceding expression in any basis in which
$B_0$ is represented by a multiple of the identity matrix.\footnote{
   Such a basis can always be achieved by a $GL(N,\C)$ gauge transformation
   $A_{xy} \to U_x A_{xy} U_y^*$.
   If the diagonal elements $A_{xx}$ are multiples of the identity
   in color space, it can be achieved simply by a {\em space-dependent}\/
   rescaling $A_{xy} \to \lambda_x A_{xy} \lambda_y^*$.
}

But the trouble is really more fundamental:
Why are we trying to minimize \reff{trace_norm}, anyway?
Minimizing \reff{trace_norm} with respect to $P$
(with $R$ and $A_1$ fixed to the variational choices)
is really no easier than minimizing the spectral radius
(as far as I can tell);
and even if it were, we wouldn't want to use the answer anyway,
because it isn't sparse.
Basically, the Hamburg condition --- even if modified to map to the right
spaces --- suffers from the same defects as Greenbaum's original condition,
without having its strongest virtues.\footnote{
   In \cite{Baker_93,Baker_LAT92} the Hamburg group actually does something
   different from what I have just described:
   Firstly, instead of $\| M_0 \|_2$ they consider
   $\| A_0^{-1} M_0 S_0^{-1} \|_2$.
   Secondly, they fix $R$ (without justifying their choice by any optimality
   condition), and then minimize \reff{trace_norm} with respect to $P$,
   imposing $RP=I$ and the Galerkin condition $A_1 = RA_0 P$
   but obviously {\em not}\/ the condition $R=P^*$.
   I do not understand the logic behind either of these elements
   of their approach.
}

The real goal, I believe, is to minimize $\rho(M_0)$
[or $\| M_0 \| _2$ or something like it],
{\em subject to some constraint on the computational complexity of the
 resulting algorithm}\/.
There seem to be two main approaches:
\begin{itemize}
   \item[1)]  {\em Geometric multigrid.}\/  Fix the sparsity pattern
     of $P$ (and thus of $A_1$), and minimize $\rho(M_0)$
     subject to this constraint.  This amounts to fixing the spatial geometry
     of the coarse grids and the interpolation.
   \item[2)]  {\em Algebraic multigrid (AMG)}\/ \cite{Ruge_87}.
     Constrain roughly the {\em number}\/ of nonzero entries in $P$ and $A_1$,
     but allow the algorithm to select the sparsity pattern
     of $P$ (and thus of $A_1$) based on some analysis of the
     matrices $A_0$ and $B_0$.
\end{itemize}
The AMG approach is potentially more powerful, but also more complicated.
In cases of very strong disorder,
such as the random-resistor problem \cite{EGS_PRL,EGS_LAT88},
AMG is probably essential;
but for gauge fields that are moderately smooth,
the geometric approach may possibly work,
and it makes sense to try it first.
The problem of minimizing $\rho(M_0)$ [or some proxy for it]
subject to a constrained sparsity pattern for $P$
needs to be examined from first principles.

Some readers may find the distinction between $V$ and $V^*$ to be
``pedantic'' or ``academic''.
I can say only that I have found this distinction to be an important
{\em clarifying}\/ principle ---
analogous to the distinction between meters and kilograms,
or between contravariant and covariant tensors in
general relativity\footnote{
   For example, the 4-velocity of a fluid is naturally a vector field,
   while the electromagnetic vector potential is naturally a {\em co}\/vector
   field (= 1-form).  In the presence of a metric tensor
   one can of course convert freely back and forth between
   vectors and covectors (``raising and lowering indices'');
   but the question is whether this operation is {\em physically natural}\/
   in the given context (that is, whether the physics of the problem at hand
   has anything to do with a spacetime metric).
   By insisting that raising-lowering operations be indicated
   {\em explicitly}\/, one facilitates such an inquiry.
   More generally, the goal is to highlight the physically relevant structures,
   and to eliminate the irrelevant ones.
   Does the problem at hand {\em really}\/ need
   all the information in a metric?
   Can it perhaps be formulated on a differentiable manifold with no additional
   structure?  Or on a manifold with only a conformal structure?
   Or on a manifold with only an affine connection?  And so forth.
}
--- which helps in sorting out the kernel of meaning
amid a mass of formulae.\footnote{
   In the multigrid literature, the distinction between $V$ and $V^*$
   can be found in \cite[Sections II.B and II.C]{Goodman-Sokal_MGMC1}
   and \cite[Section 8]{Bramble_91}.
   Possibly there are also other references of which I am unaware.
}

Other readers may object to my advocacy of a basis-independent approach
on the grounds that
``the computer must of necessity work in a particular basis''.
This is of course true, just as it is true that the article you are now
reading must of necessity be written in a particular natural language
(in this case English).
Nevertheless, the {\em content}\/ of this article should remain the same
if, for example, it is translated to Serbo-Croatian;
likewise, the {\em content}\/ of the multigrid computation should remain
the same if it is translated to another basis.
The basis has no more {\em physical}\/ relevance than does the choice
of $(x,y,z)$ axes in mechanics.
In reality, the basis has only two {\em relevant}\/ effects:
on {\em roundoff error}\/, and on {\em computational complexity}\/.
The former effect is quite easily handled by making rescalings
(or $GL(N,\C)$ gauge transformations)
to prevent the matrix elements from getting too big or too small;
this is a mere computational detail,
which does not alter the underlying algorithm.
The computational complexity is, on the other hand, a serious issue:
as mentioned above, it must be handled by imposing some constraints
on the sparsity pattern of $P$.

\bigskip

{\em Fermions.}\/
The reader is probably wondering how fermions fit into the
foregoing framework.  The answer is that I am not completely sure,
but here is my best attempt:

The underlying structure of a fermionic model seems to be
a pair of vector spaces $V$, $\bar{V}$ equipped with
a distinguished bijective map $E \colon\, V \to \bar{V}$
(here $E$ maps $\psi_i$ onto its corresponding field $\bar{\psi}_i$).\footnote{
   Equivalently, one can consider the vector space $V \oplus \bar{V}$
   equipped with a distinguished involution
   $J \colon\, V \oplus \bar{V} \to V \oplus \bar{V}$,
   where $J = \left(\!\! \begin{array}{cc}
                            0 & E^{-1} \\
                            E & 0
                         \end{array}
              \!\!\right)$.
}
The fermionic operator $A = \dslash + m$ then maps $V \to \bar{V}^*$
(think of the bilinear form $\bar{\psi}_i A_{ij} \psi_j$).

There are two main approaches to multigrid for fermions:
\begin{itemize}
   \item[1)]  Work directly with the fermionic operator $A$
     \cite{Brower_91a,Brower_91d,BenAv_90,BenAv_91,Harmatz_LAT90,BenAv_thesis,%
Lauwers_92a,Lauwers_92b,Harmatz_LAT92,Lauwers_LAT92,Lauwers_93,BenAv_93}.
   \item[2)]  Work with the ``bosonic'' operator $A^* A$ or $A A^*$
     \cite{Hulsebos_LAT90,Vink_91,Hulsebos_92a,Hulsebos_92b,Vink_LAT91,%
Kalkreuter_92c,Kalkreuter_92d,Kalkreuter_LAT92,Kalkreuter_93b}.
\end{itemize}
My personal preference is for the second approach:
it leads us back to the familiar terrain of a symmetric positive-definite
operator, for which we can confidently apply Gauss-Seidel smoothing,
the variational choices $R = P^*$ and $A_1 = P^* A_0 P$, and so forth;
the sole question is the usual one, namely the choice of the interpolation $P$.

The only trouble with the operators $A^* A$ and $A A^*$ is that,
according to my framework, they are {\em meaningless}\/!
If $A$ maps $V \to \bar{V}^*$, then $A^*$ maps $\bar{V} \to V^*$,
so the combinations $A^* A$ and $A A^*$ are meaningless.
Worse yet, there is {\em no}\/ way of making a meaningful ``squared'' map
$V \to V^*$ or $\bar{V} \to \bar{V}^*$
(remember that we want it to be positive-semidefinite,
 so it has to map a space into its dual)
out of the ingredients $A$, $A^*$, $E$, $E^*$, $E^{-1}$ and $(E^{-1})^*$!
Rather, it seems that one may consider, with equal justice,
{\em any}\/ of the operators $A^* KA$ or $ALA^*$, where
$K \colon\, \bar{V}^* \to \bar{V}$  and  $L \colon\, V^* \to V$
are symmetric and positive-definite but otherwise arbitrary.

Luckily, this formalism need be applied only to the original (fine-grid)
fermion operator, which is a Wilson or staggered fermion operator
in a {\em unitary}\/ gauge field and with a
{\em space-independent color-independent}\/ mass term.
(The coarse-grid operators, which may well involve ``dielectric'' gauge fields,
 are all bosonic.)
Therefore, it may be possible to {\em justify}\/ the standard choice
of $K$ and $L$ --- namely, the operator specified by the identity matrix
in the site-wise basis --- as being in some precise sense ``natural''.
I leave this as a open problem.

\bigskip

{\em Some concluding remarks.}\/
The foregoing comments should not be interpreted as a dismissal
of the existing literature on multigrid methods for propagators.
In fact, I believe that the two major existing approaches
--- parallel-transported multigrid
\cite{BenAv_90,BenAv_91,Harmatz_LAT90,BenAv_thesis,Lauwers_92a,Lauwers_92b,%
Harmatz_LAT92,Lauwers_LAT92,Lauwers_93,BenAv_93}
and
ground-state-projection multigrid
\cite{Hulsebos_LAT88,Hulsebos_90,Hulsebos_LAT90,Hulsebos_92a,%
Hulsebos_92b,Brower_91a,Brower_91b,Brower_91c,Brower_91d,Brower_LAT90,%
Kalkreuter_92a,Kalkreuter_92b,Kalkreuter_92c,%
Kalkreuter_92d,Kalkreuter_LAT92,Kalkreuter_93a,Kalkreuter_93b}
---
both contain valuable physical insights that are likely to play a role
in successful future algorithms.
However, I believe that these methods must be {\em rethought}\/ and
{\em reworked}\/ to bring them into conformity with the three basic
conceptual principles expounded here.

Some readers will undoubtedly feel that these conceptual principles
--- especially the distinction between $V$ and $V^*$ ---
are mere ``mathematical nit-picking'', and are irrelevant
to the practical task of designing an efficient algorithm.
I take the opposite point of view:
it seems to me that at the current stage of development,
we have already {\em too many}\/ proposals for multigrid algorithms
and their ingredients, not too few;
what we need now is to improve our {\em understanding}\/
of what we are doing, so as to lay a solid foundation for future success
in practice.
Only time will tell.

\bigskip
\bigskip

I wish to thank Radi Ben-Av, Rich Brower, Robert Edwards, Jonathan Goodman,
Thomas Kalkreuter, Gerhard Mack, Sorin Solomon and Marcus Speh
for numerous discussions over the past few years.
This research was supported in part by
National Science Foundation grant DMS--9200719.

\clearpage

\end{document}